\documentclass[A4]{jpconf}
\usepackage{graphicx}
\usepackage{color}

\begin{document}

\title{Lattice Boltzmann Model for Electronic Structure Simulations}
\author {M. Mendoza$^1$, H.J. Herrmann$^{1,3}$, S. Succi$^2$}
\address{$^1$Computational Physics for Engineering Materials, Institute for Building Materials, ETH Z{\"u}rich, Schafmattstrasse 6, HIF, CH-8093 Z{\"u}rich (Switzerland).\\ $^2$Istituto per le Applicazioni del Calcolo C.N.R., Via dei Taurini, 19 00185, Rome (Italy), and Institute for Advanced Computational Science, Harvard University,
Oxford Street, 29, MA 0213879, Cambridge, (USA).\\ $^3$Departamento de F\'{\i}sica, Universidade Federal do Cear\'a, Campus do Pici, 60455-760 Fortaleza, Cear\'a, (Brazil).}

\ead{mmendoza@ethz.ch, hans@ifb.baug.ethz.ch, succi@iac.cnr.it}

\section{Abstract}
Recently, a new connection between density functional theory and kinetic theory has been proposed. In particular, it was shown that the Kohn-Sham (KS) equations can be reformulated as a macroscopic limit of the steady-state solution of a suitable single-particle kinetic equation. By using a discrete version of this new formalism, the exchange and correlation energies of simple atoms and the geometrical configuration of the methane molecule were calculated accurately. Here, we discuss the main ideas behind the lattice kinetic approach to electronic structure computations, offer some considerations for prospective extensions, and also show additional numerical results, namely the geometrical configuration of the water molecule.

\section{Introduction}

The calculation of physical properties of interacting many-body quantum systems is one of the major challenges in chemistry and condensed matter. In principle, this requires the solution of the Schr\"odinger or Dirac equations for $3N$ spatial variables and $N$ spin electronic variables, where $N$ is the number of particles in the system. 
Usually, for atoms, the number of electrons is in the range $N \sim 1-100$, for small molecules often more than $100$, and for solids, one has around $N\sim 10^{23}$. Therefore, developing approximate models to describe these systems becomes crucial. 
In many cases, the most important goal is to calculate several measurable quantities, such as the bonding energy, polarisability, conductivity, etc., rather than the wave function itself. 
A formalism which can treat the systems mentioned with way less computational effort, while still being formally {\it exact}, is the density functional method \cite{DFTbook}, developed by Hohenberg and Kohn \cite{hohen1} and Kohn and Sham \cite{kohn1}. The Kohn-Sham approach to density functional theory allows an exact description of the interacting many-particle systems in terms of an effective non-interacting particle system. The effective potential can be shown to be completely determined by the total electron density of the interacting system, which is the reason why it is called a density functional. In particular, the ground state energy of the system is a density functional, whose exact expression is not known due to the complicated nature of the quantum many-body problem. However over the years, thanks to an intense and thoughtful work, many increasingly
more accurate approximations have continued to appear \cite{becke1, lee1}.

On an apparently very different front, fluid dynamics, the Lattice Boltzmann (LB) method has been used with considerable success to simulate a variety of physical systems \cite{statphys,graphene1,prdreldiss}. In the recent years, however, it has become increasingly apparent that the LB methodology is not confined to fluid dynamics and can indeed be extended to a variety of complex dynamical systems described by linear and non-linear partial differential equations, such as electromagnetism, quantum mechanics, Bose-Einstein condensation and many other \cite{electroLB, succi1993lattice, palpacelli2008quantum}. In a very recent paper, Mendoza {\it et al.} \cite{PRL2014} showed for the first time that the LB extends to the case of electronic structure simulations; more precisely, these authors developed and validated a lattice kinetic formulation of the Kohn-Sham equations of electrons density functional theory. In this paper, we review the main ideas and motivations behind the above formulation and discuss main challenges ahead.

\section{Electron density functional theory: Kohn-Sham equations}

The quantum mechanical behaviour of a molecule consisting of $N_I$ ions and $N_e$ electrons is described by the Schr\"odinger equation for the N-body wave function $\Psi(r_1 \dots r_{N_e}; R_1 \dots R_{N_I};t)$, that is
\begin{equation}
i \hbar \partial_t \Psi = H \Psi \quad ,
\end{equation} 
where $H$ is the N-body Hamiltonian of the system, i.e. kinetic energy plus electron-electron and electron-ion Coulomb interactions.  This equation is computationally unviable for all but the smallest molecules owing to the exponential barrier of complexity with the number of electrons \cite{kohn1999nobel}. Therefore, approximated models have been developed to study such systems.

\subsection{Kohn-Sham formulation}

In the early 60's Kohn and Hohenberg proved that the ground-state energy of a
quantum many-body system is a unique functional of the one-body electron density
$\rho(\vec{r})$ \cite{hohen1}. Starting from this basic result, shortly later Kohn and Sham, proceeded to derive a set of effective one-body time-independent Schr\"odinger equations (SE) of the form \cite{kohn1}:
\begin{equation}
\label{KSE}
H_{KS} \phi_j = E_j \phi_j \quad ,
\end{equation}  
where $\phi_j(\vec{r})$ is the wave function of the $j$-th orbital with energy $E_j$
and the KS Hamiltonian reads as follows:
\begin{equation}
H_{KS} = - \frac{\hbar^2}{2m}\nabla_j^2 + \int \frac{\rho(\vec{r}')}{|\vec{r}-\vec{r}'|} d^3r' + \sum_I Z_I \int \frac{\rho(\vec{r}')}{|\vec{R}_I-\vec{r}'|} d^3r' + V_{xc}[\rho] \quad ,
\end{equation}
where $\vec{R}_{I}$ and $Z_I$ are the coordinates and charges of the nuclei, respectively. In the above, $V_{xc}$ {\it defines} the exchange-correlation energy, the term collecting all of the unknown effects of many-body interactions and Pauli exclusion principle. Note that, by effect of the Kohn-Hohenberg theorem, this term
is guaranteed to depend only on the total electron density
\begin{equation}
\rho(\vec{r}) = \sum_j |\phi_j(\vec{r})|^2 \quad ,
\end{equation}
which represents an enormous simplification of the quantum many-body problem. If such dependence could be sorted out, the Kohn-Sham equations (KSE) would be {\it exact}.

From the mathematical point of view, the KSE's represent a set of single-particle time-independent Schr\"odinger equations, coupled only through the total density $\rho(\vec{r})$. Furthermore, the orbitals must obey the orthogonality condition
\begin{equation}
\label{ORTOG}
\int \phi_j (\vec{r}) \phi_k(\vec{r}) d\vec{r} = \delta_{jk} \quad ,
\end{equation}
which is a set of $N_e (N_e+1)/2$ global constraints. To be noted that the above orbitals are a mere computational device to perform variational minimisation of the many-body ground-state energy, with no specific physical meaning.

\section{The kinetic approach to Kohn-Sham theory}

The main observation is that, upon performing a Wick rotation, $t \rightarrow t'=-it$, the time-dependent Schr\"odinger equation is basically an one-particle diffusion-reaction equation, where the reaction term contains all the details of the potential energy, i.e.,
\begin{equation}
    \frac{\partial \psi}{\partial t'} = \frac{\hbar}{2m} \nabla^2\psi - \frac{V}{\hbar} \psi \quad ,
\end{equation}
where $\psi$ is the wave function, and $V$ is the potential energy. Since reaction-diffusion equations are known to emerge as a macroscopic limit of an underlying Boltzmann equation, it is natural to treat the wave function as a ``fluid'' and propose a (lattice) kinetic representation for the wave function, such that
\begin{equation}
\sum_p f_{p}(\vec{r};t') = \psi(\vec{r};t')  \quad ,
\end{equation}
In the above, $f_{p}(\vec{r})$, with $p=0,...,b$, is the probability of finding a ``particle'' with velocity $\vec{v}_p$ at position $\vec{r}$ and time $t$, where $b$ is the total number of discrete velocity vectors. For the moment, ``particle'' means simply a computational quasi-particle, with no physical implications for the existence of an underlying kinetic theory behind the Schr\"odinger representation. Note that assuming an arbitrary initial condition for the wave function $\psi$, one can expand it using a basis of orthogonal orbitals, 
\begin{equation}
    \psi(\vec{r},t') = \sum_{k} a_{k} \phi_k (\vec{r}) \exp(-E_k t'/\hbar) \quad ,
\end{equation}  
such that after some time, it will be $\psi(\vec{r},t') \simeq a_0 \phi_0(\vec{r})\exp(-E_0 t'/\hbar)$, which upon normalisation leads to the ground state solution of the time-independent Schr\"odinger equation,
\begin{equation}
    E_0 \phi_0 = -\frac{\hbar^2}{2m} \nabla^2 \phi_0 + V\phi_0 \quad .
\end{equation}
The distribution function $f_p$ evolves according to the lattice Boltzmann equation (time step made unity for simplicity),
\begin{equation}
\label{LBeq}
f_{p}(\vec{r}+\vec{v}_p ;t+1) = f_{p}(\vec{r};t) - \omega (f_{p}-f_{p}^e) + V_{p} \quad , 
\end{equation}
where $\omega$ is a typical relaxation frequency which controls the quantum diffusivity $D=\hbar/m$, $f_{p}^e$ is a suitable local equilibrium ensuring mass conservation, and $V_{p}$ is a suitable source term representing the particle loss/gain associated with the potential energy scattering. In lattice units $\Delta x = \Delta t = 1$, one has:
\begin{equation}
D = c_s^2 (1/\omega - 1/2) \quad ,
\end{equation}
where $c_s$ is the lattice sound speed. Thus, by changing $\omega$, one has a handle on the effective mass of the electrons. The explicit expression for the equilibrium distribution is as follows:
\begin{equation}
f_{p}^e = \xi_p \psi \quad , 
\end{equation}
where $\xi_p = (1+(D-c_s^2)q_p)$, with $w_p$ being suitable lattice weights normalised to unity, and $q_p = (v_p^2 - 3c_s^2)/2 c_s^4$ the second order lattice Hermite polynomial. The scattering source $V_p$ reads as follows:  
\begin{equation}
V_{p} =  \chi_p \frac{V}{\hbar} \psi \quad ,
\end{equation}
where $\chi_p = w_p[1 - (v_p^2 - 3c_s^2)/2c_s^2)]$ ensures that the first and second order moments are zero, namely $\sum_p f_p \vec{v}_p = 0$ and $\sum_p f_p \vec{v}_p \vec{v}_p = 0$, so that momentum and energy of the wave function are unaffected. 

In order to solve the KS system, we need a distribution function for each $j$th-orbital, i.e. $f_p \rightarrow f_{jp}$ and consequently $\psi \rightarrow \psi_j$, and the potential energy $V$ contains the electron-electron and ion-electron interactions, plus the exchange-correlation potential. Thus, the corresponding electronic Lattice Boltzmann (LB) equation reads as follows:
\begin{equation}
\label{KSLB}
f_{jp}(\vec{r}+\vec{v}_p ;t+1) = f_{jp}(\vec{r};t) - \omega (f_{jp}-f_{jp}^e) + V_{jp} + W_{jp} \quad , 
\end{equation}
where, $W_{jp}$ is the source term in charge of securing the orthogonality constraints (\ref{ORTOG}). The explicit expressions for the equilibrium distributions and scattering potentials are as follows:
\begin{equation}
f_{jp}^e = \xi_p \psi_j, \quad {\rm and}\quad  V_{jp} =  \chi_p \frac{V}{\hbar} \psi_j \quad ,
\end{equation}
and the orthogonalisation potential takes the form
\begin{equation}
W_{jp} =  -\xi_p \omega \sum_{k<j} \Lambda_{jk} \psi_k \quad ,
\end{equation}
where $\Lambda_{jk} = \langle \psi_j|\psi_k\rangle/\langle\psi_k|\psi_k\rangle$ is the cosine of the angle between orbitals $j$ and $k$ in Hilbert space. Clearly, once two orbitals are mutually orthogonal, $\Lambda_{jk}=0$, so that they no longer contribute to the potential $W_{jp}$. 

In order to understand the role of the orthogonalisation potential we will introduce the following derivation. Let us first assume that all orbitals are initialised with the same wave function, $\psi_j(\vec{r}) = \psi (\vec{r})$, and we do not consider any orthogonalisation potential in equation (\ref{KSLB}), i.e. $W_{jp} = 0$. According to our previous explanation, the wave functions can be written as expansions in the orthogonal basis $\phi_k (\vec{r})$,
\begin{equation}
    \psi_j(\vec{r},t') = \sum_p f_{jp} = \sum_{k} a_{kj} \phi_k (\vec{r}) \exp(-E_k t'/\hbar) \quad .
\end{equation}
where $a_{kj} = \langle \phi_k|\psi_j\rangle$ are projection coefficients. After some time, the only leading term in the sum will be the ground state orbital, $\phi_0$, such that $\psi_j (\vec{r},t') \simeq a_{0j} \phi_0(\vec{r}) \exp(-E_0 t'/\hbar)$. At this stage, all orbitals in our lattice kinetic approach will reach the same ground state. Let us now subtract from the wave function $\psi_1$, the wave function $\psi_0$, such that 
\begin{equation}
    \psi_1(\vec{r},t') = \sum_{k} a_{k1} \phi_k (\vec{r}) \exp(-E_k t'/\hbar) - \Lambda_{10} \psi_0({\vec{r},t'}) \quad .
\end{equation}
This equation can be rewritten explicitly as,
\begin{equation}
    \psi_1(\vec{r},t') = \sum_{k} \left( \langle \phi_k|\psi_1\rangle - \frac{\langle \psi_1|\psi_0\rangle}{\langle \psi_0|\psi_0\rangle} \langle \phi_k|\psi_0 \rangle \right) \phi_k (\vec{r}) \exp(-E_k t'/\hbar) \quad .
\end{equation}
The evolution equation for the orbital $\psi_0$ implies that it will converge eventually to the ground state $\phi_0$. Therefore, the first term in the sum, for $k = 0$, vanishes and the remaining contributions to the sum are
\begin{equation}
    \psi_1(\vec{r},t') = \sum_{k>0} a_{k1} \phi_k (\vec{r}) \exp(-E_k t'/\hbar) \quad .
\end{equation}
Thus, the orbital $\psi_1$ will eventually converge to the first excited state $\phi_1$. Repeating this procedure for each wave function, one can write  
\begin{equation}
    \psi_j(\vec{r},t') = \sum_{k} \left( \langle \phi_k|\psi_j\rangle -  \sum_{l < j}\frac{\langle \psi_j|\psi_l\rangle}{\langle \psi_l|\psi_l\rangle} \langle \phi_k|\psi_l \rangle \right) \phi_k (\vec{r}) \exp(-E_k t'/\hbar) \quad ,
\end{equation}
or 
\begin{equation}\label{ortho_de}
    \psi_j(\vec{r},t') = \sum_{k} a_{kj} \phi_k (\vec{r}) \exp(-E_k t'/\hbar) - \sum_{l<j} \Lambda_{jl} \psi_l(\vec{r},t') \quad ,
\end{equation}
and show that each wave function will eventually converge, upon normalisation, to the respective KS orbital, i.e. $\psi_j \rightarrow \phi_j$. In order to include this procedure into the evolution of the wave functions, we need to force the $j$th wave function to converge to
\begin{equation}\label{ortho_de}
    \psi_j^*(\vec{r},t') = \psi_j(\vec{r},t') - \sum_{l<j} \Lambda_{jl} \psi_l(\vec{r},t') = \sum_{p} f_{jp} - \sum_{l<j} \Lambda_{jl} \psi_l(\vec{r},t') \quad .
\end{equation}
This can be done by replacing in the equilibrium distribution $\psi_j$ by $\psi_j^*$, obtaining the following lattice kinetic equation,
\begin{equation}
\label{KSLB-2}
f_{jp}(\vec{r}+\vec{v}_p ;t+1) = f_{jp}(\vec{r};t) - \omega (f_{jp}-\xi_p \psi_j^*) + V_{jp} \quad .
\end{equation}
By reorganising terms, this equation leads to
\begin{equation}
\label{KSLB-2}
f_{jp}(\vec{r}+\vec{v}_p ;t+1) = f_{jp}(\vec{r};t) - \omega (f_{jp}-\xi_p \psi_j) + V_{jp} - \xi_p \omega \sum_{l<j} \Lambda_{jl} \psi_l(\vec{r},t')\quad , 
\end{equation}
which is precisely our previous definition of the lattice kinetic equation with the orthogonalisation potential. Note that after some time, we have $\psi_j \rightarrow \psi_j^* \rightarrow \phi_j$. This process is a kind of time-dependent Gram-Schmidt procedure. Indeed, one can also evolve first the wave function $\psi_0$, and once the ground state is found, evolve $\psi_1$, and so on. However, this will lead to longer computational times to achieve the total electronic ground state. The scaling properties and convergency speed of the orthogonalisation procedure described here will be a subject of future research.

\subsection{Concurrent Dynamics}

The above electronic LB only applies to ground-state calculations, where the
electronic distribution is systematically adjusted to the Born-Oppenheimer surface
defined by the actual positions of the ions, $R_I(t)$, these latter being obtained
by a standard Molecular Dynamics integration. The total energy is thus evolving in (imaginary) time until the total ground state value is attained. This is the standard Born-Oppenheimer (BO) scenario.

The KS-LB can be extended to the concurrent-dynamic (CD) scenario, whereby the electrons and the ions are moved simultaneously, according to the celebrated
Car-Parrinello picture \cite{carpari}. In this scenario, energy owes to be conserved, and consequently the irreversible relaxation operator in Eq.~(\ref{KSLB}) needs to be augmented with a potential, $U$, such that density and momentum are conserved, while energy is supplied to (removed from) the system so as to keep it constant.

The corresponding LB equation takes then the form:
\begin{equation}
\label{KSLBCD}
f_{jp}(\vec{r}+\vec{v}_p h ;t+h) = 
f_{jp}(\vec{r}-\vec{v}_p h;t-h) + 
2h(U_{jp} + iV_{jp} + iW_{jp}) \quad ,
\end{equation}
where $h$ is the time-step. The specific expression of $U_{jp}$ is as follows:
\begin{equation}
U_{jp} =  \kappa w_p q_p E_j \phi_j \quad ,
\end{equation}
$\kappa$ being an adjustable parameter in charge of keeping the energy at $E_j$.
To be noted that the algorithm is now a two-step time-centered scheme, in compliance with the requirement of energy conservation. This is the analogue, in kinetic language, of Car-Parrinello's inertial term in the electron equation, i.e. a second order time derivative to comply with the overarching electron-ion Lagrangian. The fact that the present LB scheme operates successfully also in the CD regime hints at the existence of an overarching kinetic Lagrangian. At the time of this writing, however, such Lagrangian remains to be found.  

\section{Results}

\begin{figure}
  \centering
  \includegraphics[width=0.5\columnwidth]{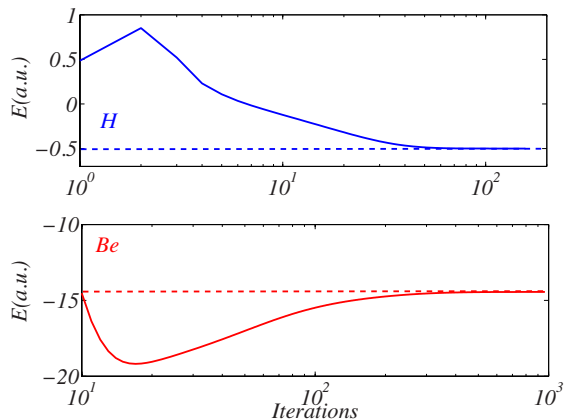}
  \caption{Convergence of the total energy for the atoms H and Be as a function of the number of iterations performed by our model. The dashed-line corresponds to the final value, $E = 0.500$ a.u. ($1$ a.u. $\simeq 27.2$ eV) and $E = 14.45$ a.u., for the hydrogen and beryllium atoms, respectively. For Beryllium, we found an error of $1\%$ with respect to the expected value in Refs.~\cite{Be1,Be2}, while for hydrogen the error is of the order of $0.1\%$. We have taken $\tau_K = \delta t$.}
\label{fig6}
\end{figure}

\begin{table}
  \centering
  \begin{tabular}{|c|c|c|c|c|c|c|}\hline
    Atom & ${\cal V}_x$ & Exp.~${\cal V}_x$ & ${\cal V}_c$ & Exp.~${\cal V}_c$ & Time & Iterations\\ \hline
    H & $-0.310$ & $-0.310$ & $0$ & $0$ & $9$ sec & $158$ \\ \hline
    He & $-1.025$ & $-1.025$ & $-0.044$ & $-0.044$ & $1$ min & $906$ \\ \hline
    Be & $-2.658$ & $-2.658$ & $-0.095$ & $-0.095$ & $3$ min & $965$ \\ \hline 
    B & $-3.728$ & $-3.728$ & $-0.128$ & $-0.128$  & $29$ min & $2704$ \\ \hline
    C & $-5.032$ & $-5.032$ & $-0.161$ & $-0.161$  & $33$ min & $2305$ \\ \hline
  \end{tabular}
  \caption{Exchange-Correlation energies for H, He, Be, B, and C in atomic units (a.u.). Computational time and the number of iterations performed by the model to reach the ground state are also shown. The expected values of exchange, ${\cal V}_x$, and correlation, ${\cal V}_c$, energies are taken from Refs.~\cite{becke1, lee1}. The simulations have run on an Intel Core i7 at 2.3 GHz. We have taken $\tau_K = \delta t$.} \label{xcvalues}  
\end{table}

In order to show the validity of our approach, we present the calculations of the exchange and correlation energies for different atoms, H, He, Be, B, and C. For H, Be, B, and C, one can find more details about the calculations and parameter values in Ref.~\cite{PRL2014}. For those cases, we have used lattice sizes of $32^3$, $40^3$, $56^3$, and $58^3$, respectively. For He, which has not been reported before using our model, we have taken a system size of $36^3$, Bohr radius $a_0 = 6.5$, $\hbar/m = 1$, and $\tau = 1$ (all values are in numerical units). We have also used fourth order in the Hermite expansion for the equilibrium distribution and source term (for details about the Hermite expansion, see Ref.~\cite{PRL2014}). The measured values for the exchange and correlation energies are given in Table~\ref{xcvalues}, together with the computational time and number of iterations. Note that the reported number of iterations might exceed the one of existing iterative methods. However, as we shall show shortly, each of our iterations takes typically much less computational time, thus making our method competitive. In Fig.~\ref{fig6}, we see that the convergence of the total energy, for H and Be, to the ground state is very fast in the beginning and slows down when it approaches the exact solution. The differences between the obtained and expected exchange and correlation energies are of the order of $1 \%$. The simulations are stopped when the energy of the orbitals presents changes of less than $10^{-5} \%$ between two subsequent steps.

\begin{figure}
  \centering
  \includegraphics[width=0.5\columnwidth]{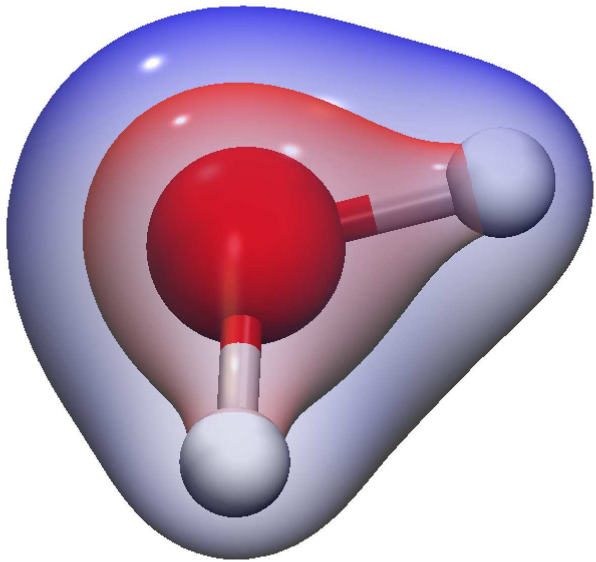}
  \caption{Water molecule, H$_2$O. The blue and red isosurfaces
    denote low and high electronic density, respectively. Using our
    model, we have obtained for the angles between bonds, $104.4$
    degrees, and a O-H bond distance of $0.95$ \AA.}
\label{fig_h2o}
\end{figure}
As an additional application, we build the water molecule from scratch, by using the {\it bare} Coulomb potential.  For this simulation, we use a lattice size of $74^3$, $\hbar/m = 1$, $a_0 = 10.5$ cells, and ${\cal N} = 5$, and place the oxygen atom at the center of the lattice. The hydrogen atoms are located randomly in space, and we let the system evolve to the ground configuration. After $4$ hours ($10447$ iterations), we achieved the configuration shown in Fig.~\ref{fig_h2o}. The angle, $104.4$ degrees, and the bond distance of $0.95$ \AA\; are in excellent agreement with expected and experimental values \cite{water1,water2}.

\begin{figure}
  \centering
  \includegraphics[width=0.5\columnwidth]{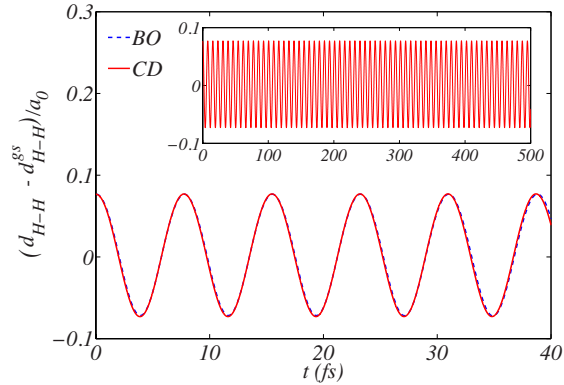}
  \caption{Vibrational mode of the diatomic molecule H$_2$. Here BO and CD denote Born-Oppenheimer and Concurrent Dynamics, respectively. The inset shows that up to $500$ fs, the amplitude of the oscillations does not show any appreciable decay in time. Here, $a_0$ is the Bohr radius, $d_{H-H}$ the distance between H atoms and $d_{H-H}^{gs}$ the ground state distance.}
  \label{fig5}
\end{figure}

Finally, we compare BO and CD versions of our kinetic scheme against each other. Here, we assume that the BO dynamics provides the correct results, and will be used to validate the CD. To this purpose, we excite the H$_2$ molecule and let it vibrate in its first mode. We have used a lattice size of $24^3$. Further details about the values of the model parameters can be found in Ref.~\cite{PRL2014}. For the BO dynamics we use the equation of motion for the ions and, at each time step, we calculate the ground state of the electronic density using Eq.~(\ref{KSLB}). On the other hand, for CD we evolve simultaneously the ions and the electronic orbitals, using the discrete version of Eq.~(\ref{KSLBCD}). Since in CD there is no relaxation process to the ground state, we must use a smaller time step such that electrons can accommodate according to the new position of the ions at each time step. The time step of the BO molecular dynamics is set to $\Delta t_{BO} = 0.09$ fs, while for CD, $\Delta t_{CD} = 0.0014$ fs. Although the time step in CD is smaller by nearly two orders of magnitude, it runs $17$ times faster (taking $17$ sec to simulate one fs) than BO dynamics. In fig.~\ref{fig5}, we report the first vibrational mode of the hydrogen molecule with BO and CD after $40$ fs, from which one can appreciate that the trajectories of CD are pretty close to the respective BO ones. In the inset of this figure, we can also see that after $500$ fs, the amplitude of the oscillations remains constant, without any appreciable loss of energy. The oscillation frequency of the vibrational mode obtained by the simulation is $4206$ cm$^{-1}$, which departs by $1\%$ from the experimental value of $4161$ cm$^{-1}$ \cite{H2freq}.

Interestingly, the LB simulations do resolve the full Coulomb interactions, without making any use of soft pseudo-potentials \cite{PRL2014}. This is possibly due to the high-order of spatial accuracy achieved by the source terms. The amount of computational memory that our model uses can be roughly estimated as $M_T = 8 (36 + 2N_O) V$ in bytes, where $N_O$ is the number of orbitals and $V$ the lattice size. For instance, for the water molecule the total memory used is about $150$ MB, and by using a standard computer (or a single node of a cluster) with $8$ GB of RAM, one could study about $8$ of these molecules. This small system size implies that parallel computing is a crucial step in order to study larger system sizes, and the above relation for the memory requirements provides a reasonable estimate of the maximum system size that one can simulate at a given level of computing power.  

Our simulations have been performed using open boundary conditions. However, the use of periodic boundary conditions is also possible at the price of introducing corrections due to Ewald summation for the electric potential. Indeed, using periodic boundary conditions can lead to more stable simulations, since the distribution functions do not need to be extrapolated at the boundaries, and therefore, their dynamics or relaxation to ground state follows the correct equations holding in the bulk.

\section{Prospective developments}

The kinetic representation discussed in this paper lends itself to a number of prospective developments. First, one should note that such representation is not restricted to the time-independent Kohn-Shan formulation. Indeed, in the recent past Runge and Gross have extended Kohn-Hohenberg's theorem to the case of time-dependent potentials \cite{runge1984density}. In this case, the KS formalism still applies, although in real-time form, i.e:
\begin{equation}
\label{TDKS}
i \hbar \partial_t \phi_j(\vec{r};t) = H_{KS}(t) \phi_j(\vec{r};t) \quad ,
\end{equation}
where the KS Hamiltonian now contains a time-dependent potential, $H_{KS} = K + V(\vec{r};t)$.

The above equations can be integrated using consolidated quantum 
LB schemes \cite{succi1993lattice}, exporting the very same upgrades described
in this paper. Another direction, possibly even closer in spirit to kinetic 
theory, is the so-called {\it density-current} formulation of TDDFT, in 
which the hamiltonian depends not only on the electron density, but also
on the electron current:
\begin{equation}
\label{TDKS}
\vec{J}(\vec{r};t) = \sum_j \vec{v}_j |\phi_j(\vec{r};t)|^2
\end{equation}
where $\vec{v}_j \equiv i \frac{\hbar}{m} \nabla_j$ is the velocity operator. 

The density and current functionals evolve according to the continuity equation:
\begin{equation}
\partial_t \rho + \nabla \cdot \vec{J} = 0
\end{equation}
which provides $\rho(\vec{r};t)$ once the current functional $\vec{J}=\vec{J}(\rho(\vec{r};t))$ is prescribed. Of course, the latter is not known exactly, but sensible approximations have been developed in the specialised literature, e.g the Vignale-Kohn functional \cite{vignale}, which could be implemented in the corresponding electronic LB exactly along the lines described in this paper.

Finally, one may also envisage a more radical strategy, whereby the energy functional would be expressed in terms of the Boltzmann distribution $f$ itself rather than by its lower-order kinetic moments, such as density, current, and momentum flux. The prospective Boltzmann functional would symbolically read as
\begin{equation}
E = \int \mathcal{E}[f] d\vec{r} d\vec{v}
\end{equation}
where the double integration extends over the six-dimensional phase space.

What would one gain out of such functional? 
Possibly, the non-perturbative incorporation of inhomogeneity effects at {\it all orders}. 
Indeed, the Boltzmann distribution splits naturally into an equilibrium and non-equilibrium 
component, $f=f^e + f^{ne}$, the former depending parametrically only on invariants, density 
and momentum, while the latter, by definition, collects the effects of space-time inhomogeneity at all orders. 
Of course, such functional would make the difference only at the level where further 
moments on top of current-density, are required. 

\section{Conclusions}
Summarising, we have presented the main ideas behind the kinetic formulation of electronic density functional theory, along with the ensuing implementation within the Lattice Boltzmann formalism. We have described in detail the effects of the orthogonalisation potential and shown how it contributes to find the system of orthogonal KS orbitals. In order to show the validity of the lattice Boltzmann approach for KS, we have shown numerical data for the computation of the exchange and correlation energies for simple atoms, finding excellent agreement with the previous literature. 
The results of numerical simulations for the first vibrational mode of the hydrogen molecule are also presented. The prospects, including future extensions to time-dependent DFT, look exciting.

\section*{Acknowledgements} 

Illuminating discussions with K. Bravaya, D. Coker and E. Kaxiras are kindly acknowledged. 
Financial support from the European Research Council (ERC) Advanced Grant 319968-FlowCCS is kindly acknowledged. 

\section*{References}

\bibliographystyle{iopart-num}
\bibliography{report}

\providecommand{\newblock}{}
\begin{thebibliography}{10}
\expandafter\ifx\csname url\endcsname\relax
  \def\url#1{{\tt #1}}\fi
\expandafter\ifx\csname urlprefix\endcsname\relax\def\urlprefix{URL }\fi
\providecommand{\eprint}[2][]{\url{#2}}

\bibitem{DFTbook}
Gross E and Dreizler R 1995 {\em Density Functional Theory\/} NATO ASI Series:
  Physics (Springer)

\bibitem{hohen1}
Hohenberg P and Kohn W 1964 {\em Phys. Rev.\/} {\bf 136}(3B) B864--B871

\bibitem{kohn1}
Kohn W and Sham L~J 1965 {\em Phys. Rev.\/} {\bf 140}(4A) A1133--A1138

\bibitem{becke1}
Becke A~D 1988 {\em Phys. Rev. A\/} {\bf 38}(6) 3098--3100

\bibitem{lee1}
Lee C, Yang W and Parr R~G 1988 {\em Phys. Rev. B\/} {\bf 37}(2) 785--789

\bibitem{statphys}
Succi S 2008 {\em The European Physical Journal B-Condensed Matter and Complex
  Systems\/} {\bf 64} 471--479

\bibitem{graphene1}
Mendoza M, Herrmann H and Succi S 2013 {\em Scientific Reports\/} {\bf 3}

\bibitem{prdreldiss}
Mendoza M, Karlin I, Succi S and Herrmann H 2013 {\em Physical Review D\/} {\bf
  87} 065027

\bibitem{electroLB}
Mendoza M and Mu\~noz J~D 2010 {\em Phys. Rev. E\/} {\bf 82}(5) 056708

\bibitem{succi1993lattice}
Succi S and Benzi R 1993 {\em Physica D: Nonlinear Phenomena\/} {\bf 69}
  327--332

\bibitem{palpacelli2008quantum}
Palpacelli S and Succi S 2008 {\em Physical Review E\/} {\bf 77} 066708

\bibitem{PRL2014}
Mendoza M, Succi S and Herrmann H~J 2014 {\em Phys. Rev. Lett.\/} {\bf 113}(9)
  096402

\bibitem{kohn1999nobel}
Kohn W 1999 {\em Reviews of Modern Physics\/} {\bf 71} 1253--1266

\bibitem{carpari}
Car R and Parrinello M 1985 {\em Phys. Rev. Lett.\/} {\bf 55}(22) 2471--2474

\bibitem{Be1}
Mori-S{\'a}nchez P, Wu Q and Yang W 2005 {\em The Journal of chemical
  physics\/} {\bf 123} 062204

\bibitem{Be2}
Hirata S, Ivanov S, Grabowski I and Bartlett R~J 2002 {\em The Journal of
  chemical physics\/} {\bf 116} 6468--6481

\bibitem{water1}
Sprik M, Hutter J and Parrinello M 1996 {\em The Journal of chemical physics\/}
  {\bf 105} 1142--1152

\bibitem{water2}
Benedict W, Gailar N and Plyler E~K 1956 {\em The Journal of Chemical
  Physics\/} {\bf 24} 1139--1165

\bibitem{H2freq}
Stoicheff B~P 1957 {\em Canadian Journal of Physics\/} {\bf 35} 730--741

\bibitem{runge1984density}
Runge E and Gross E~K 1984 {\em Physical Review Letters\/} {\bf 52} 997

\bibitem{vignale}
Vignale G and Kohn W 1996 {\em Phys. Rev. Lett.\/} {\bf 77}(10) 2037--2040

\end{thebibliography}

\end{document}